\documentclass[a4paper,12pt]{article}
\pdfoutput=1 %arXiv fully supports and automatically recognizes PDFLaTeX. You can ensure pdflatex processing by setting the flag \pdfoutput=1 within the first 5 lines of the preamble of the main .tex file. You should not need any other special flag.
\usepackage{graphicx}% Include figure files
\usepackage{bm}% bold math
\def\be{\begin{equation}}
\def\ee{\end{equation}}
\def\bea{\begin{eqnarray}}
\def\eea{\end{eqnarray}}
\usepackage{latexsym}
\usepackage{dcolumn}
\usepackage{amsmath}
\usepackage{epsfig,amssymb,euscript,mathrsfs}
\usepackage{array,calc,epsfig}

\usepackage[noadjust]{cite}

\usepackage{chngcntr}

\usepackage[pdfpagelabels]{hyperref}

\usepackage{color}

\def\be{\begin{equation}}
\def\ee{\end{equation}}
\def\ba{\begin{eqnarray}}
\def\ea{\end{eqnarray}}
\def\nb{\nonumber}

\counterwithin*{equation}{section}

\def\({\left (}
\def\){\right )}

\let\benonumb\[
\let\eenonumb\]

\def\[{\left [}
\def\]{\right ]}

\let\oldlgraf\{ %rinomino la parentesi graffa con un altro nome, così posso usare \{ per la ridefinzione del comando
\renewcommand{\{}{\left \oldlgraf}

\let\oldrgraf\}
\renewcommand{\}}{\right \oldrgraf}

\usepackage{xfrac} %per frazioni storte (\sfrac{}{})
\usepackage{float} %per \begin{figure}[H]
\usepackage{subfig}

\newcommand{\eqnref}[1]{(\ref{#1})}

\newlength\dlf  % Define a new measure, dlf

\usepackage{booktabs}
\usepackage{comment}
\usepackage{cases} %per scrivere sistemi di equazioni labellate
\usepackage{empheq}
\usepackage{cancel}

\begin{document}
\baselineskip=15.5pt
\pagestyle{plain}
\setcounter{page}{1}
\newfont{\namefont}{cmr10}
\newfont{\addfont}{cmti7 scaled 1440}
\newfont{\boldmathfont}{cmbx10}
\newfont{\headfontb}{cmbx10 scaled 1728}
\renewcommand{\theequation}{{\rm\thesection.\arabic{equation}}}
\renewcommand{\thefootnote}{\arabic{footnote}}

\vspace{1cm}

\begin{titlepage}
\vskip 2cm
\begin{center}
{\Large{\bf Higher Order Corrections to the Hagedorn Temperature at Strong Coupling
}}
\end{center}

\vskip 10pt
\begin{center}
Francesco Bigazzi$^{a}$, Tommaso Canneti$^{a,b,c}$, Aldo L. Cotrone$^{a,b}$
\end{center}
\vskip 10pt
\begin{center}
\vspace{0.2cm}
\textit {$^a$ INFN, Sezione di Firenze; Via G. Sansone 1; I-50019 Sesto Fiorentino (Firenze), Italy.
}\\
\textit{$^b$ Dipartimento di Fisica e Astronomia, Universit\'a di Firenze; Via G. Sansone 1;\\ I-50019 Sesto Fiorentino (Firenze), Italy.
}\\
\textit{$^c$ Institute for Advanced Study, School of Natural Sciences%
\protect\\ Princeton, NJ 08540, USA.
}
\vskip 20pt
{\small{
bigazzi@fi.infn.it, canneti@fi.infn.it, cotrone@fi.infn.it}
}

\end{center}

\vspace{25pt}

\begin{center}
 \textbf{Abstract}
\end{center}

\noindent 

We propose a general formula for higher order corrections to the value of the Hagedorn temperature of a class of holographic confining gauge theories in the strong coupling expansion.
Inspired by recent proposals in the literature, the formula combines the sigma-model string expansion with an effective approach. In particular, it includes the sigma-model contributions to the Hagedorn temperature at next-to-next-to leading order, which are computed in full generality. 
For ${\cal N}=4$ SYM on $S^3$ our result agrees with numerical estimates with excellent precision. We use the general formula to predict the value of the Hagedorn temperature for ABJM on $S^2$ and for the dual of purely RR global $AdS_3$.

\end{titlepage}

\newpage
\tableofcontents

\section{Introduction}

In recent months, the determination of the Hagedorn temperature $T_H$  \cite{Hagedorn:1968jf} in string backgrounds has found renewed interest. In \cite{Bigazzi:2022gal,Urbach:2023npi,Bigazzi:2023oqm}, the authors address the problem in supergravity solutions dual to confining gauge theories, both with a world-sheet method and with an effective approach inspired by the Horowitz-Polchinski construction \cite{Horowitz:1997jc}. The latter has been also applied to exact backgrounds as $AdS_5 \times S^5$ \cite{Urbach:2022xzw}. This made possible to infer the Hagedorn temperature for the dual $\mathcal{N}=4$ SYM theory on $S^3$ up to next-to-leading order (NLO) in the holographic limit. The outcome is in agreement with the numerical analysis of \cite{Harmark:2017yrv, Harmark:2018red, Harmark:2021qma} based on integrability and quantum spectral curve methods. In any case, an analytical computation of the next-to-next-to-leading order (NNLO) correction was missing.

Recently the authors of \cite{Ekhammar:2023glu} proposed a method to extend the calculation in the effective approach to higher orders. Apart from a sigma-model contribution which cannot be captured by the approach, they calculate the NNLO and NNNLO terms of $T_H$ for $\mathcal{N}=4$ SYM theory on $S^3$ in the planar limit. The starting point is the effective action for a scalar field $\chi$ corresponding to the winding mode of a string sitting at the center of the dual $AdS$ background. The string wraps the compact Euclidean time direction once and the corresponding winding mode becomes light near the Hagedorn temperature. The Hagedorn temperature is determined requiring that, neglecting the backreaction on the background, the linearized equation of motion for the scalar field admits a normalizable solution.  To NLO in the holographic limit, solving the scalar equation of motion near the $AdS$ center
amounts to solve the eigenvalue problem for an harmonic oscillator in flat space \cite{maldanotes}. At NNLO the effective Hamiltonian of this equivalent quantum model gets a further interaction term. The corresponding correction to the ground state energy can be then computed using perturbation theory \cite{Ekhammar:2023glu}. This allows to infer part of the NNLO corrections to the Hagedorn temperature. Indeed, going beyond the NLO, the higher-order corrections to the world-sheet sigma model, which are not accounted for by the effective approach, also come into play. The related NNLO contribution, $\Delta c$ in the notation of \cite{Ekhammar:2023glu}, remained undetermined. In order to get insights on its possible analytic expression, the authors considered, as a benchmark, a type IIB string on an exactly solvable pp-wave background, obtained as a Penrose limit of  the original $AdS_5\times S^5$ one. The value of $\Delta c$ extracted from this construction is however not in agreement with the one inferred from the numerical results from integrability. The authors of \cite{Ekhammar:2023glu} noticed that the correct coefficient would have been $\Delta c= - 4 \log2$. They also anticipated a forthcoming analysis \cite{toappear} of a similar construction involving global $AdS_4$, reporting that in that case $\Delta c = -3\log2$ would fit the numerical results on $T_H$. They thus conjectured $\Delta c/\log2$ to be related to the dimension of the dual CFT. 

In this paper we present a first principle sigma model derivation of $\Delta c$ in a large class of holographic confining theories. In the global $AdS$ cases, our results confirm the above expectations. They also imply that, in general non-$AdS$ cases, $\Delta c/\log2$ is not equal to the dimensionality of the space where the dual field theories live.  

The proper world-sheet approach which allows to find the value of $\Delta c$ has been developed in \cite{Bigazzi:2023oqm} for a specific case. It relies on the expansion of the world-sheet sigma model (in the near-Hagedorn regime) up to second order in quantum fluctuations around a reference configuration describing the winding string sitting at the center of the background. Due to the curvature of the latter, some of the bosonic fluctuations get a non-trivial mass and they can be decomposed as a sum of zero and non-zero modes. As observed in \cite{Bigazzi:2023oqm}, the effective approach takes into account just the former ones. Since the Hagedorn temperature at NLO depends only on the bosonic massive zero modes, the two methods give the same result to this order of approximation. As we will see in a moment, this observation holds in full generality. At NNLO, the quadratic contribution of the bosonic and fermionic non-zero modes is well-understood in the world-sheet approach, while it is outside of reach of the effective approach. On the other hand, the quartic contribution of the zero modes is not encoded in the (quadratic) sigma-model, while it is captured by the effective theory as described in  \cite{Ekhammar:2023glu}.

The aim of this article is thus to use the interplay of the two methods as a tool to infer a general expression for the Hagedorn temperature of a class of confining gauge theories with a dual string theory description, up to NNNLO in the holographic limit. 

The holographic backgrounds dual to the low energy regime of finite temperature confining gauge theories display string frame metrics whose asymptotic form can be generically written as
\be
ds^2 \approx 2\pi\alpha' T_s \left(1+ \frac{r^2}{l^2}\right) (dt^2 + \eta_{ij}dx^i dx^j) + dr^2 + r^2 d\Omega_{d-1}^2 + ds^2_{{\cal M}}\,.
\label{metricb}
\ee
Here $r$ is the holographic radius, with $r\rightarrow0$ corresponding to the IR of the dual gauge theory. Moreover $T_s$ is the confining string tension, $l$ is the curvature radius, $i,j = 1,...p$ and ${\cal M}$ is a compact $(9-p-d)$-dimensional transverse space. Finally, the Euclidean time coordinate is compactified on a circle of length $\beta=1/T$ where $T$ is the gauge theory temperature. Global $AdS_{D}\times{\cal M}_{10-D}$ backgrounds are included in this class setting $p=0$ and $d=D-1$; the dual field theories are ``confining'' in the sense explained by \cite{Witten:1998zw}. The backgrounds will also contain (possibly running) dilatons and Ramond-Ramond fluxes over the transverse compact space. We assume that, if Kalb-Ramond fluxes are present, they become trivial at the center of the background. 

A closed string, sitting at $r=0$ and at a point in the transverse space, and winding once along the compact time circle (at zero momentum) is described by the embeddings $x^{N}=x^N(\tau,\sigma)$ where, in particular
\be\label{config}
x^0\equiv t = \frac{\beta}{2\pi} \sigma + \xi^0 (\tau,\sigma)\,.
\ee
The first term ensures periodicity under $\sigma\mapsto\sigma+2\pi$. At the Hagedorn temperature the ground state of the string is massless \cite{Sathiapalan:1986db,Kogan:1987jd,OBrien:1987kzw,Atick:1988si}, i.e. $M^2 = - \eta^{ij}p_i p_j =0$, where $p^i$ are the momenta along the $x^i$ directions. As a result, starting from the Polyakov action on the above background, one realizes that, to linear order in the fluctuations, the $d$ scalar fields $y^I(\tau,\sigma)$ where $y^I y^I = r^2$, satisfy Klein-Gordon equations with masses
\be\label{genmu}
\mu= \frac{\beta_H}{2\pi}\frac{\sqrt{2\pi \alpha' T_s}}{l}\,.
\ee
Thus, our proposal for the inverse Hagedorn temperature $\beta_H=1/T_H$ is
\be \label{genproposal}
\frac{T_s}{2} \, \beta_H^2 = 2\pi \[ \, \Delta(\mu) + \Delta {\cal E} \, \] \, ,
\ee
where 
$\Delta(\mu)$ is the (finite) zero point energy of the world-sheet sigma model expanded up to second order in the bosonic mass parameter $\mu$,
and $\Delta {\cal E}$ contains the NNLO and NNNLO contributions coming from the effective approach.
For theories dual to global $AdS_{D}$-type backgrounds, explored in section \ref{sec:examples}, $T_s=1/2\pi\alpha'$ and $l$ is equal to the $AdS$ radius $R_{AdS}$, so that $\mu=\beta_H/(2\pi R_{AdS})$.

Let us give a few details on the function $\Delta(\mu)$. The metrics (\ref{metricb}) display a shrinking $(d-1)$-cycle at $r=0$ and the string worldsheet theory contains $d$ massive bosons $y^I$ of mass $\mu$, $8-d$ massless bosons and $8$ massive antiperiodic fermions with masses $f_i \mu\,,i=1,...,8$. 

Let us now further specialize our analysis to a class of holographic theories with no running dilaton. In those cases, cancelation of Weyl anomaly implies that the sum of the bosonic squared masses has to be equal to that of the fermionic ones (see e.g.~\cite{Drukker:2000ep,Bigazzi:2004ze,Gautason:2021vfc}), so that 
\be
\sum_{i=1}^{8}f_i^2 = d\,.
\label{condm}
\ee
The function $\Delta(\mu)$ is just the zero-point energy of this system
\be
\Delta(\mu)=- (8-d) \sum_{n=1}^{\infty} n - \frac{d}{2}\mu - d\sum_{n=1}^{\infty}\sqrt{n^2+\mu^2} + \sum_{i=1}^8\sum_{n=1}^{\infty}\sqrt{\left(n-\frac12\right)^2+ f_i^2\mu^2}\,.
\label{Delta}
\ee
The ${\cal O}(\mu)$ term is due to the zero modes of the massive bosonic fields. Expanding to quadratic order in $\mu$ one gets\footnote{This formula can be trivially generalized to the case of unequal world-sheet scalar masses $a_i \mu, i=1,...,8$, see (\ref{formul}).}
\be
\label{deltaexp}
\Delta(\mu)= 1 -\frac{d}{2}\mu + d\, \mu^2 \log{2} + {\cal O}(\mu^4)\,,
\ee
which, crucially, does not depend on the $f_i$ due to the condition (\ref{condm}). 

To NLO in the strong coupling expansion, {\it i.e.} to first order in $\mu$, combining (\ref{deltaexp}), (\ref{genproposal}) and (\ref{genmu}), our result reproduces the general formula for the Hagedorn temperature found in~\cite{Urbach:2023npi}. This is coherent with the observation that, to this order, both the effective approach and the sigma-model one account for just the bosonic zero modes. Moreover, to this order, conditions like eq. (\ref{condm}) do not come into play. Hence, to NLO, our result holds also for holographic confining theories with running dilatons, like the so-called WYM model with D4s wrapped on a circle \cite{Witten:1998zw}, where $d=2$,
and Maldacena-Nu\~nez \cite{Maldacena:2000yy} models where $d=3$. Another confining model with constant dilaton, but running Kalb-Ramond form $B_2$, is the Klebanov-Strassler \cite{Klebanov:2000hb} one, for $d=3$.

Crucially, for holographic confining theories with no running dilaton and trivial $B_2$ fluxes, writing $\Delta(\mu) = 1 -(d/2)\mu - \mu^2 \Delta c + {\cal O}(\mu^4)$ we get, from eq. (\ref{deltaexp}), the general prediction
\be \label{gendeltac}
\Delta c = - d \log 2 \,,
\ee
for the NNLO sigma-model contribution.

This fills the gap in the analysis in \cite{Ekhammar:2023glu}. 
Indeed, for $\mathcal{N}=4$ SYM theory on $S^3$, whose dual background features a global $AdS_5$ factor, we have $d=4$ and thus $\Delta c = - 4 \log 2$, which agrees with the numerical expectations from integrability. On the other hand, notice that, from (\ref{Delta}), above the ${\cal O}(\mu)$-order the expansion of $\Delta$ is in even powers of $\mu$. This means that the quadratic sigma-model corrections do not enter at NNNL order, which instead receives contributions from second-order perturbation theory in the effective approach.
However, in principle, products of couples of zero modes with couples of non-zero modes in the quartic string action could affect the result at this order. For the time being, we have no control on these corrections. A-posteriori, the precise matching of the string results with the field theory ones shows that these corrections are absent (or very small) in $AdS$ backgrounds. But, strictly speaking, the assumption that the world-sheet does not contribute at NNNLO is still a conjecture.
Finally, if the background is exact in $\alpha'$, one could expect that there are no other corrections to this order.
For generic backgrounds dual to confining theories the latter is not true, so the analysis above is expected to be exact only up to NNLO. 

In more generality, formula (\ref{gendeltac}) gives $\Delta c = -(D-1) \log 2$ for any gauge theory dual to global $AdS_D\times{\cal M}_{10-D}$ backgrounds with constant dilaton and no $B_2$ fluxes. Hence, for the ABJM theory \cite{Aharony:2008ug} on $S^2$, whose dual description exhibits a global $AdS_4$, we get $\Delta c = -3\log 2$ in agreement with the expectations in \cite{toappear} reported in \cite{Ekhammar:2023glu}.

Moreover, formula (\ref{gendeltac}) directly provides the sigma model contribution to the NNLO term in the Hagedorn temperature of confining theories (within the special class we are focusing on) obtained as the low energy dynamics of wrapped D-branes on $S^{d-1}$. We thus have, for example, $\Delta c = -2 \log 2$ for the case of D3s compactified on a circle \cite{Witten:1998zw}.

A way to go beyond NNNLO is to derive a full-fledged world-sheet description. This goes beyond the purposes of this paper, which is actually based on the interplay between complementary approaches. We defer such a development, as well as the cases of running dilaton and NS-NS fluxes, to future works.

This work is structured as follows. In section 2, we will provide a general explicit expression for our formula (\ref{genproposal}) to NNNLO for any gauge theory dual to global $AdS_{d+1}$ backgrounds with constant dilaton and no $B_2$ fluxes. In section 3 we will apply this formula to three cases. First to $\mathcal{N}=4$ SYM on $S^3$, reproducing the numerical results in \cite{Harmark:2017yrv, Harmark:2018red, Harmark:2021qma}. Then, to the ABJM case, providing an analytic prediction for the Hagedorn temperature to be checked with hopefully forthcoming numerical results. Finally to the case where the gravity description is provided by $AdS_3$ with only Ramond-Ramond fluxes. In appendix \ref{app:ppwave} we will briefly review the computation of the Hagedorn temperature in the plane-wave limit of the backgrounds discussed in the paper, signaling a mismatch with the results from (\ref{genproposal}). This underlines how the plane-wave limit cannot be used as a guide to deduce the higher-order corrections to the Hagedorn temperature of the full theory, as it has been done in \cite{Ekhammar:2023glu}.

\section{Global $AdS$ backgrounds}
\label{sec:exact}

In this section, we aim to give a detailed derivation of the Hagedorn temperature for $d$-dimensional CFTs compactified on $S^{(d-1)}$, up to NNNLO in the holographic limit. Therefore, we focus on string backgrounds whose metric is of the factorized form global-$AdS_{d+1} \times {\cal M}_{9-d}$. If these are exact solutions in $\alpha'$, we have just to deal with corrections coming from the world-sheet sigma model and higher-order perturbations to the effective Hamiltonian of the winding mode. From an effective point of view, the method which we are going to describe is just a generalization of the one in \cite{Ekhammar:2023glu}. Our contribution completes the treatment implementing the world-sheet data derived in a complementary approach. In the following, we will fix the $AdS$ radius $R_{AdS}$ to one and we will work with dimensionless inverse temperature $\beta/R_{AdS} \mapsto \beta$ and dimensionless string scale $\alpha'/R_{AdS}^2 \mapsto \alpha'$ in order to declutter the equations.

Let us start from the effective approach. As we have recalled in the Introduction, the near Hagedorn behavior of the string model can be captured by an effective field theory for a complex scalar field $\chi$ winding once the thermal circle. Reducing the theory on the compact directions, the Hagedorn singularity happens when the winding mode $\chi$ becomes massless. This is the picture adopted in \cite{Urbach:2023npi, Urbach:2022xzw} for the NLO computation of $T_H$. 

Starting from scratch, global $AdS_{d+1}$ can be described by the metric
\be
ds^2 = (1+R^2) d\tau^2 + \frac{dR^2}{1+R^2} + R^2 d\Omega^2_{d-1} \, .
\ee
The $d$-dimensional effective action which describes the infrared dynamics of a scalar field
\be
\chi = \chi(R)
\ee
winding once the thermal direction of this background is
\be
S_{\chi} \approx \beta \hspace{-4pt} \int \hspace{-4pt} dR \, R^{d-1} \{ g^{pq} \,  \partial_p \chi^* \, \partial_q \chi + m^2_{eff}(R) \, \chi^* \chi\}\,,
\ee
where
\be
m^2_{eff}(R) = (1+R^2)\left(\frac{\beta}{2\pi\alpha'} \right)^2 - \frac{2}{\alpha'} \,. 
\ee
The equation for the mode $\chi$ is thus
\be \label{geneom}
-\frac{1}{R^{d-1}} \partial_R\left(R^{d-1} (1+R^2)\partial_R\right) \chi(R) + m^2_{eff}(R) \, \chi(R)=0\,.
\ee

At quadratic order (i.e.~dropping the $R^2$ term in parenthesis in the derivative term), the equation is the one of a $d$-dimensional harmonic oscillator with frequency $\omega=\beta_H/2\pi\alpha'$ and energy levels $E_n=\omega(n+d/2)$.
In other words, the ``unperturbed'' problem takes the form of
\be
-\frac12 \chi''(R) - \frac12 (d-1) \frac1R \chi'(R) + \frac12 \omega^2 R^2 \chi(R) = \omega \(n+\frac{d}{2}\) \chi(R) \, , \quad \omega = \frac{\beta_H}{2 \pi \alpha'} \, ,
\ee
once given the normalizability condition
\be \label{impnormcond}
\omega \(n+\frac{d}{2}\) = \frac12 \[ \frac{2}{\alpha'} - \omega^2  \] \, .
\ee
As the integer $n$ varies, the solutions of the above equation are thus the well-known eigenfunctions of zero-angular momentum
\be
\chi_n(R) = \alpha_n e^{-\frac{\omega}{2} R^2} L_{\sfrac{n}{2}}^{\sfrac{d}{2}-1} (\omega R^2) \, ,
\ee
where $ L_{\sfrac{n}{2}}^{\sfrac{d}{2}-1}$
are the associated Laguerre polynomials, and $\alpha_n$ are normalization constants.
The normalization is calculated with the scalar product
\be 
\langle \chi_n | \chi_m \rangle = \int_0^{\infty} dR \, R^{d-1} \, \chi_n^* \, \chi_m\,.
\ee
In this way, the relevant eigenfunctions for this method read
\be
\chi_0(R) = \[\frac12 \, \omega^{-\sfrac{d}{2}} \, \Gamma\(\frac{d}{2}\)\]^{-\sfrac{1}{2}} \hspace{-4pt} e^{-\frac\omega2 R^2}
\ee
and
\be
\chi_4(R) = \[\omega^{-\sfrac{d}{2}} \, \Gamma\(2+\frac{d}{2}\)\]^{-\sfrac{1}{2}} \hspace{-4pt} e^{-\frac\omega2 R^2} \(\omega^2 R^4 - (d+2) \omega R^2 + \frac14 d (d+2)\) \, .
\ee

Now, we can reintroduce the $R^2$ term in parenthesis in the derivative term in \eqnref{geneom} as the perturbed Hamiltonian
\be
\Delta H = - \frac12 \[(d+1) R \frac{\partial}{\partial R} + R^2 \frac{\partial^2}{\partial R^2} \] \, .
\ee
Therefore, one can immediately derive the first order correction to the ground state energy as
\be
\Delta E^{(1)} = \left \langle \chi_0 | \Delta H| \chi_0 \right \rangle = \frac{d \, (d+2)}{8}
\ee
and, limiting ourselves to the leading contribution, the second order one as
\be
\Delta E^{(2)} \approx \frac{\left | \left \langle \chi_0 | \Delta H| \chi_4 \right \rangle \right |^2}{- 4 \, \omega} = - \frac{d(d+2)\pi\alpha'}{16 \, \beta} \, .
\ee
Introducing also the NNLO world-sheet contribution $\Delta c$ as in \cite{Ekhammar:2023glu}, the condition \eqnref{impnormcond} for the ground state is thus corrected as
\be \label{NNNLO}
\frac{d}{2} \frac{\beta}{2\pi\alpha'} + \Delta E^{(1)} + \Delta E^{(2)} = \frac12 \[ \frac{2}{\alpha'} - \frac{\beta^2}{2\pi^2\alpha'} \Delta c - \(\frac{\beta}{2\pi\alpha'}\)^2  \] \, .
\ee
Crucially, as we have shown in equation \eqnref{gendeltac}, our world-sheet method gives
\be
\Delta c = - d \log 2  \, .
\ee
Therefore, exploiting the relation between $\mu = \beta_H/2\pi$ and $\alpha'$ given by the NLO version of \eqnref{NNNLO}, we can rephrase the latter up to NNNLO as
\be \label{finalimplicitrel}
\frac{T_s}{2} \beta_H^2 = 2\pi \[ 1 - \frac{d}{2} \, \mu + \(d \, \log 2 - \frac{d(d+2)}{16} \) \mu^2 + (1-4d)\frac{d(d+2)}{16} \frac{\mu^3}{8} \] \, , \quad T_s = \frac{1}{2\pi\alpha'}  \, .
\ee
Of course, this is an implicit equation for $\beta_H=1/T_H$. Solving for it and defining the coupling $g=1/4\pi\alpha'$, our final result reads
\be \label{finalgenresult}
\boxed{
T_H = \sqrt{\frac{g}{2\pi}} + \frac{d}{8\pi} + \frac{d^2+d-8d\log2}{32\sqrt{2}\pi^{\sfrac{3}{2}}\sqrt{g}} + \frac{4d^3+7d^2-2d}{1024 \pi^2 g} + \mathcal{O}(g^{-\sfrac{3}{2}}) 
} \, .
\ee
Remember that the above is a dimensionless quantity. To get the physical result, just divide by $R_{AdS}$.

%%%%%%%%%%%%%%%%%%%%%%%%%%%%%%%%%%%%%%%%%%%%%%%%%%%

%%%%%%%%%%%%%%%%%%%%%%%%%%%%%%%%%%%%%%%%%%%%%%%%%%%
\section{Examples}
\label{sec:examples}

The goal of this section is to make explicit the value of the Hagedorn temperature $T_H$ for some relevant cases, up to NNNLO in the holographic limit.
%%%%%%%%%%%%%%%%%%%%%%%%%%%%%%%%%%%%%%%%%%%%%%%%%%%
\subsection{$\mathcal{N}=4$ SYM theory on $S^3$}

Let us consider $\mathcal{N}=4$ SYM on $S^3$ at strong coupling.
In this case $d=4$.  
As explained in the Introduction between formulas (\ref{config}) and (\ref{genmu}), the worldsheet theory has four massless and four massive bosonic modes with masses $\mu= \beta_H/(2\pi R_{AdS})$.
Thus, relation (\ref{condm}) implies that the eight fermionic modes have masses equal to $\mu_f =\mu \sqrt{d}/{2\sqrt{2}}=\mu /{\sqrt{2}}$.
From our general formula \eqnref{finalimplicitrel} we recover
\be
\frac{\beta_H^2}{4\pi\alpha'} = 2\pi - 2 \, \beta_H + \frac{\beta_H^2}{2 \pi} \, 4 \log 2 - 6\pi \alpha' +\frac{3\pi^2\alpha'^2}{\beta_H}  \, ,
\ee
where the last two terms on the right hand side have been first computed in \cite{Ekhammar:2023glu}, while the third one is new. 
Resolving for $T_H$ we get (as a special case of our formula \eqref{finalgenresult})
\begin{align}
T_H &= \frac{1}{\beta_H} = \sqrt{\frac{g}{2\pi}} + \frac{1}{2\pi} + \frac{5 - 8 \log 2}{8\pi\sqrt{2\pi}\sqrt{g}} + \frac{45}{128 \pi^2 g} + \mathcal{O}(g^{-3/2}) \\
& \approx 0.3989\,\sqrt{g} + 0.159 \, - \frac{0.0087}{\sqrt{g}} + \frac{0.036}{g} + \mathcal{O}(g^{-3/2}) \nb \, .
\end{align}
To NLO the outcome was already found in \cite{Urbach:2022xzw}. This result is in very good agreement with the numerical results from integrability \cite{Harmark:2021qma,Ekhammar:2023glu}, within the declared uncertainties.

%%%%%%%%%%%%%%%%%%%%%%%%%%%%%%%%%%%%%%%%%%%%%%%%%%%
\subsection{ABJM theory on $S^2$}
\label{sec:abjm}

Here we report the value of the Hagedorn temperature in the ABJM theory \cite{Aharony:2008ug} up to NNNL order at strong coupling. The background is global-$AdS_4 \times CP^3$. 
In this case $d=3$, so that
\be \label{finalabjm}
\frac{\beta_H^2}{4\pi\alpha'} = 2\pi- \frac32 \beta_H + \frac{\beta_H^2}{2\pi} 3  \log{2}  -\frac{15}{4}\pi \alpha' +\frac{15}{8}\frac{\pi^2 \alpha'^2}{\beta_H}\,.
\ee
Notice that our formula \eqnref{gendeltac} gives
\be
\Delta c = - 3 \log 2 \, .
\ee
This result is in perfect agreement with the expectation in \cite{toappear} anticipated in \cite{Ekhammar:2023glu}.
Finally, 
the Hagedorn temperature reads
\be  
T_H = \sqrt{\frac{g}{2\pi}} + \frac{3}{8\pi} + \frac{3(1-2\log{2})}{8\sqrt{2}\pi^{3/2} \sqrt{g}} + \frac{165}{1024\pi^2 g}  + \mathcal{O}(g^{-\sfrac{3}{2}}) \,.
\ee

%%%%%%%%%%%%%%%%%%%%%%%%%%%%%%%%%%%%%%%%%%%%%%%%
\subsection{Pure RR global-$AdS_3$ case}
\label{sec:ads3}

We move on to write the result for the purely RR $AdS_3$ case in global coordinates.
In this case we have two massive bosonic modes and $d=2$.
Thus
\be
\frac{\beta_H^2}{4\pi\alpha'} = 2\pi - \beta_H  + \frac{\beta_H^2}{\pi} \log{2}  -2 \pi \alpha' +\frac{\pi^2 \alpha'^2}{\beta_H} \,.
\ee
Note that $\Delta c = - 2 \log 2$ and the final result is
\be
T_H = \sqrt{\frac{g}{2\pi}} + \frac{1}{4\pi} + \frac{3-8\log{2}}{16\sqrt{2}\pi^{3/2} \sqrt{g}} + \frac{7}{128\pi^2 g}  + \mathcal{O}(g^{-\sfrac{3}{2}}) \,.
\ee

%%%%%%%%%%%%%%%%%%%%%%%%%%%%%%%%
\vskip 15pt \centerline{\bf Acknowledgments} \vskip 10pt 

\noindent 
We thank Federico Castellani, Wolfgang M\"uck, Giulio Pettini and Erez Y. Urbach for comments and very helpful discussions.

%%%%%%%%%%%%%%%%%%%%%%%%%%%%%%%%%%%%%%%%%%%%%%%%%%%%%%%%%%

\appendix

%%%%%%%%%%%%%%%%%%%%%%%%%%%%%%%%%%%%%%%%%%%%%%%%%%%%%%%
\section{Comparison with pp-wave results}
\label{app:ppwave}

In this appendix we review the extraction of the Hagedorn temperature in the plane-wave regime of the $AdS_5 \times S^5$ and $AdS_4 \times CP^3$ backgrounds, pointing out that its value is different from the one in (\ref{genproposal}) beyond NL order.
This shows that the pp-wave limit is not suitable to obtain information on the Hagedorn temperature of the full theory (as proposed in \cite{Ekhammar:2023glu}). 

Consider a quadratic world-sheet theory of eight bosons with masses $m_{B,i}^2 = a_i^2 \mu^2, i=1,...,8$ and eight fermions with masses $m_{F,i}^2 = b_i^2 \mu^2, i=1,...,8$.
Absence of anomalies imply the mass matching condition
\begin{equation}\label{matching}
\sum_{i=1}^8 a_i^2  = \sum_{i=1}^8 b_i^2 \,.
\end{equation}
Let us define the function
\begin{equation}
\Delta(s,\mu) = - \sum_{i=1}^8\sum_{n=1}^{\infty} (n^2+a_i^2 \mu^2)^{-s/2} -\frac12 \sum_{i=1}^8 (a_i \mu)^{-s} + \sum_{i=1}^8\sum_{n=1}^{\infty} \left[\left(n- \frac12 \right)^2 + b_i^2 \mu^2  \right]^{-s/2}\,.
\end{equation}
This function can be employed to define the Hagedorn temperature along the lines of \cite{Bigazzi:2023oqm}.
Let us consider its expansion around $\mu=0$ up to quadratic order
\begin{equation}
\Delta(s,\mu) =  -8(2-2^s)\zeta(s)  -\frac12 \sum_{i=1}^8 (a_i \mu)^{-s} + \sum_{i=1}^8 (a_i^2 \mu^2)s\zeta(s+2) -\sum_{i=1}^8 (b_i^2 \mu^2)s 2^{s+1}\zeta(s+2) + ... \,.
\end{equation}
The last two terms can be combined due to the mass matching condition (\ref{matching}).
Using the Laurent expansion of the zeta function $\zeta(s+2) = 1/(s+1) +$regular, in the limit $s \rightarrow -1$ one has ($\zeta(-1)=-1/12$)
\begin{equation}\label{formul}
\Delta(\mu) = 1 -\frac12 \sum_{i=1}^8 (a_i \mu) + \log{2} \sum_{i=1}^8 (a_i^2 \mu^2) + ... \,.
\end{equation}

For a plane-wave metric of the form
\be 
ds^2 = 2 dx^+ dx^- - \mu^2 \sum_{i=1}^8 (a_i x_i)^2 dx^+ dx^+ + dx^idx_i\,,
\ee
the Hagedorn temperature $T_H=1/\beta_H$ is defined by (see e.g.~formula (28) of \cite{Bigazzi:2003jk})
\begin{equation}
\frac{\beta_H^2}{4\pi \alpha'}=2\pi \Delta\left(\frac{\mu \beta_H}{2\sqrt{2}\pi} \right)\,.
\end{equation}

%%%%%%%%%%%%%%%%%%%%%%%%%%%%%%%%%%%%%%%%%%%%
\subsection{${\cal N}=4$ SYM}
As a first example, let us consider $AdS_5 \times S^5$.
In the pp-wave limit, the string theory dual to ${\cal N}=4$ SYM reduces to a quadratic world-sheet theory with eight massive bosons of the same mass $a_i \mu=f, i=1,...,8$.
Thus
\begin{equation}
\Delta(m) = 1 - 4m + 8m^2\log{2} + ... \,,
\end{equation}
and
\begin{equation}
\frac{\beta_H^2}{4\pi \alpha'}=2\pi -2\sqrt{2} f \beta_H + \frac{2\log{2}}{\pi} f^2 \beta_H^2 +...\,.
\end{equation}
This is exactly the expression in (3.16) of \cite{Ekhammar:2023glu}.
It provides the correct NL term of $T_H$ of the full theory,  but it fails to give the correct (i.e.~matching the numerical calculation from integrability) contribution at NNL order due to a missing factor of two in the last term. 

%%%%%%%%%%%%%%%%%%%%%%%%%%%%%%%%%%%%%%%%%%%%
\subsection{ABJM}

The pp-wave limit in this case gives a quadratic world-sheet theory with four massive bosons of mass $\mu^2=1/2$ and four massive bosons of mass $\mu^2=1/8$~\cite{Nishioka:2008gz}.
Thus
\be  
\Delta = 1- \frac{3}{\sqrt{2}} + \frac{5 \log{2}}{2}\,, 
\ee
and
\be
\frac{\beta_H^2}{4\pi\alpha'} = 2\pi- \frac32 \beta_H + \frac{\beta_H^2}{8\pi} \, 5 \log{2}\,.
\ee
While the leading and NL terms coincide with the ones from (\ref{genproposal}), the NNL term is different from the one in (\ref{finalabjm}).

\newpage
%%%%%%%%%%%%%%%%%%%%%%%%%%%%%%%%%%%%%%%%%%%%%%

\end{document}